\documentstyle[12pt]{article}
\def\beq{\begin{equation}}
\def\eeq{\end{equation}}
\def\bea{\begin{eqnarray}}
\def\eea{\end{eqnarray}}
\def\ba{\begin{array}}
\def\ea{\end{array}}
\def\lrb{\left(}
\def\rrb{\right)}
\def\lsb{\left[}
\def\rsb{\right]}
\def\lcb{\left\{}
\def\rcb{\right\}}
\def\nn{\nonumber}
\def\pl{\partial}
\def\p{\prime}
\def\dlb{\lsb\hskip -0.6mm\lsb}
\def\drb{\rsb\hskip -0.6mm\rsb} 
\def\one{1\hskip -1.1mm{\rm l}}
\begin{document}
\begin{flushright}
IMSc/2000/03/09
\end{flushright}
\begin{center}
{\bf AN INTRODUCTION TO QUANTUM ALGEBRAS \\
AND THEIR APPLICATIONS} 
\footnote{Article based on {\em Narayani, N. and Kadayam S. Sankaran Memorial 
Lectures} delivered by the author at the Department of Mathematics, Ramakrishna 
Mission Vivekananda College, Mylapore, Chennai, on 23 \& 27 January 1999} 

\bigskip

Ramaswamy JAGANATHAN \\
{\em The Institute of Mathematical Sciences \\
C.I.T. Campus, Tharamani, Chennai - 600 113, India \\
E-Mail}\,:{\tt jagan@imsc.ernet.in} 
\end{center}

\vspace{0.5cm}

\noindent 
{\bf Abstract}\,: A very elementary introduction to quantum algebras is presented 
and a few examples of their physical applications are mentioned. 

\bigskip

\noindent 
I shall give a very elementary introduction to the topic of quantum algebras 
and mention a few physical applications.  Quantum algebras, or quantum groups, 
extend the domain of classical group theory and constitute a new and growing 
field of mathematics with vast potential for applications in physics.  In fact, 
the origins of quantum groups lie in physics\,: in the studies on the behaviour 
of integrable systems in quantum field theory and statistical mechanics, using 
the quantum inverse scattering method, by Sklyanin, Kulish, Reshetikhin, 
Takhtajan, and Faddeev in the 1980s.  Mathematical abstraction from the 
observations on the common features of these systems led to the definition of 
the concept of a quantum group and the studies of Drinfeld, Jimbo, Manin, 
Woronowicz, Connes, Wess, Zumino, Macfarlane, Biedenharn, ..., revealed 
many aspects of quantum groups from different mathematical and physical points 
of view.  I shall neither go into the history of the developements of the 
various significant concepts nor give any specific references to their origins.  
I shall cite at the end some books and articles which would give these 
details and lead to the vast literature on the subject of quantum groups 
and their applications.     

Let us consider a two dimensional classical vector space whose elements can be 
written as 
\beq
\lrb \ba{c} x \\ y \ea \rrb\,,
\eeq
where the coordinates $x$ and $y$ are real variables and commute with each 
other, {\em i.e.}, 
\beq
xy = yx\,.
\eeq
In the space of functions $\lcb f(x,y) \rcb$ defined in the above vector space 
the partial derivative operations $\frac{\pl}{\pl x}$ and $\frac{\pl}{\pl y}$ 
and the operations of multiplications by $x$ and $y$ satisfy the differential 
calculus 
\bea 
[x\,,\,y]\ =\ 0\,, \quad  
\lsb \frac{\pl}{\pl x}\,,\,\frac{\pl}{\pl y} \rsb & = & 0\,, \quad   
\lsb \frac{\pl}{\pl x}\,,\,y \rsb\ =\ 0\,, \quad 
\lsb \frac{\pl}{\pl y}\,,\,x \rsb\ =\ 0\,, \nn \\ 
\lsb \frac{\pl}{\pl x}\,,\,x \rsb & = & 1\,, \quad 
\lsb \frac{\pl}{\pl y}\,,\,y \rsb\ =\ 1\,, 
\label{cdc}
\eea
where the commutator bracket $[\,\,,\,]$ is defined by 
\beq
[A\,,\,B] = AB-BA\,.
\eeq  
Let us make a linear transformation of the vector $\lrb \ba{c} x \\ y \ea \rrb$ 
as 
\beq
\lrb \ba{c} x^\p \\ y^\p \ea \rrb 
 = M \lrb \ba{c} x \\ y \ea \rrb  
 = \lrb \ba{cc} a & b \\ c & d \ea \rrb 
   \lrb \ba{c} x \\ y \ea \rrb 
\label{sl2tr}
\eeq
where the entries of the matrix $M$ are real and satisfy the condition 
\beq
ad - bc = \mbox{det}\,M = 1\,, 
\eeq
or in other words the transformation~(\ref{sl2tr}) is an element of the group  
$SL(2,R)$. Then, the new coordinates $x^\p$ and $y^\p$ and partial derivatives 
with respect to them, namely $\frac{\pl}{\pl x^\p}$ and $\frac{\pl}{\pl y^\p}$, 
also satisfy the same relations as~(\ref{cdc}).  This is easily checked by noting 
\beq
\lrb \ba{c} 
     \frac{\pl}{\pl x^\p} \\ 
                          \\
     \frac{\pl}{\pl y^\p} 
\ea \rrb = \lrb \ba{rrr} d & & -c \\ & & \\ -b & & a \ea \rrb 
\lrb \ba{c} 
     \frac{\pl}{\pl x} \\ 
                       \\ 
    \frac{\pl}{\pl y} 
\ea \rrb = \widetilde{M^{-1}} 
\lrb \ba{c} 
     \frac{\pl}{\pl x} \\ 
                       \\
     \frac{\pl}{\pl y} 
\ea \rrb\,, 
\eeq
where $\widetilde{A}$ means the transpose of the matrix $A$.  We say that the 
differential calculus on the two dimensional $(x,y)$-plane is covariant under 
the group $SL(2,R)$. 

In nature all physical systems are quantum mechanical.  But, the quantum 
mechanical behaviour is generally revealed only at the microscopic (molecular 
and deeper) level.  At the macroscopic level of everyday experience quantum 
physics becomes classical physics  as an approximation.  So, only classical 
physics was discovered first and observations of failure of classical physics 
at the atomic level led to the discovery of quantum physics in the 20th 
century.  This passage from classical physics to quantum physics can be 
mathematically described as a process of deformation of the classical physics 
wherein the commuting classical observables of a physical system are replaced 
by noncommuting hermitian operators.  This process is characterized by a very 
small deformation parameter known as the Planck constant $\hbar$ and, roughly 
speaking, in the limit $\hbar \longrightarrow 0$ classical physics is 
recovered from the structure of quantum physics.  

In analogy with the process of quantizing the classical physics let us now 
quantize the classical vector space to get a quantum vector space by assuming 
that the coordinates do not commute with each other at any point.  Let us model 
the noncommutativity of the coordinates $X$ and $Y$ of a two dimensional 
quantum plane by 
\beq
XY = qYX\,, 
\label{XY}
\eeq
where $q$ is the deformation parameter which we shall consider, in general, 
to be any nonzero complex number.  Note that in the limit $q \longrightarrow 1$ 
the noncommuting quantum coordinates $X$ and $Y$ become commuting classical 
coordinates.  To be specific, let us choose 
\beq
q = e^{i\theta}\,.
\eeq
It is easy to find an example of such noncommuting variables.  Let $T_\alpha$ 
and $G_{\theta/\alpha}$ be operators acting on functions of a single real 
variable $x$ such that 
\beq
T_\alpha\psi(x) = \psi(x+\alpha)\,, \quad 
G_{\theta/\alpha}\psi(x) = e^{i\theta x/\alpha}\psi(x)\,.
\eeq
Then, for any $\psi(x)$, 
\beq
T_\alpha G_{\theta/\alpha}\psi(x) 
   = e^{i\theta(x+\alpha)/\alpha}\psi(x+\alpha) 
   = e^{i\theta}G_{\theta/\alpha}T_\alpha\psi(x)\,, 
\eeq
for a given fixed value of $\theta$.  Thus, with the variation of $\alpha$, 
$T_\alpha$ and $G_{\theta/\alpha}$ become noncommuting variables obeying the 
relation 
\beq
T_\alpha G_{\theta/\alpha} = e^{i\theta}G_{\theta/\alpha}T_\alpha\,,
\eeq
with fixed value for $\theta$.  

Now the interesting questions are\,: 
\begin{itemize}
\item Is it possible to define a differential calculus on the two dimensional 
$(X,Y)$-plane?   
\item If so, will it be covariant under some generalization of the classical 
group $SL(2)$? 
\end{itemize}  
The answers are yes, yes!  First let us understand how one can give a meaning 
to partial derivatives with respect to $X$ and $Y$.  These have to operate on 
the space of functions $\lcb f(X,Y) \rcb$ which we shall consider to be 
polynomials in $X$ and $Y$.  We can write $f(X,Y) = \sum_{m,n}f_{mn}X^mY^n$ 
since any polynomial in  $X$ and $Y$, with coefficients commuting with $X$ and 
$Y$, can be brought to this form using the commutation relation~(\ref{XY}).  
Then, if we take formally 
\beq
\frac{\pl}{\pl X}X^m = mX^{m-1}\,, \quad 
\frac{\pl}{\pl Y}Y^n = nY^{n-1}\,, 
\eeq
we would have a differential calculus in the $(X,Y)$-plane, as desired, once we 
prescribe consistently the remaining commutation relations between $X$, $Y$, 
$\frac{\pl}{\pl X}$ and $\frac{\pl}{\pl Y}$.  

Without going into any further details, let me state the complete set of 
commutation relations defining the differential calculus in the $(X,Y)$-plane 
(with $q = e^{i\theta}$)\,: 
\bea
&  & XY = qYX\,, \quad 
\frac{\pl}{\pl X}\frac{\pl}{\pl Y} 
   = q^{-1}\frac{\pl}{\pl Y}\frac{\pl}{\pl X}\,, \quad 
\frac{\pl}{\pl X}Y = qY\frac{\pl}{\pl X}\,, \quad 
\frac{\pl}{\pl Y}X = qX\frac{\pl}{\pl Y}\,, \nn \\ 
&  & \frac{\pl}{\pl X}X - q^2X\frac{\pl}{\pl X} 
         = 1+(q^2-1)Y\frac{\pl}{\pl Y}\,, \quad 
\frac{\pl}{\pl Y}Y - q^2Y\frac{\pl}{\pl Y} = 1\,. 
\label{XYrel}
\eea
This noncommutative differential calculus on the two-dimensional quantum plane 
is seen to be covariant under the transformations 
\bea
\lrb \ba{c} X^\p \\ Y^\p \ea \rrb  
 & = & T \lrb \ba{c} X \\ Y \ea \rrb  
 = \lrb \ba{cc} A & B \\ C & D \ea \rrb 
   \lrb \ba{c} X \\ Y \ea \rrb \nn \\  
\lrb \ba{c} 
     \frac{\pl}{\pl X^\p} \\ 
                          \\
     \frac{\pl}{\pl Y^\p} 
\ea \rrb & = & \widetilde{T^{-1}} 
\lrb \ba{c} 
     \frac{\pl}{\pl X} \\ 
                       \\
     \frac{\pl}{\pl Y} 
\ea \rrb = \lrb \ba{ccc} D & & -qC \\ 
                               & &     \\ 
                      -q^{-1}B & & A 
                    \ea \rrb 
\lrb \ba{c} 
     \frac{\pl}{\pl X} \\ 
                       \\
     \frac{\pl}{\pl Y} 
\ea \rrb\,,  
\label{Ttr}
\eea 
provided
\beq
A,B,C,\ \mbox{and}\ D\ \mbox{commute with}\ X\ 
          \mbox{and}\ Y\,, 
\eeq
\bea
   &   & AB = qBA\,, \quad CD = qDC\,, \quad 
         AC = qCA\,, \quad BD = qDB\,, \nn \\ 
   &   & BC =  CB\,, \quad 
         AD-DA = \lrb q-q^{-1} \rrb BC\,, 
\label{Trel}
\eea
and
\beq
AD-qBC = \mbox{det}_qT = 1\,.
\eeq
In other words 
$X^\p, Y^\p, \frac{\pl}{\pl X^\p},\ \mbox{and}\ \frac{\pl}{\pl Y^\p}$ 
defined by~(\ref{Ttr}) satisfy the relations obtained from~(\ref{XYrel}) by 
just replacing $X$ and $Y$ by $X^\p$ and $Y^\p$ respectively.  Note that 
$\mbox{det}_qT$ defined in~(\ref{Trel}) commutes with all the matrix elements 
of $T$.  Verify that 
\beq
T^{-1} = \lrb \ba{cc} D & -q^{-1}B \\ -qC & A \ea \rrb\ 
\eeq
is such that 
\beq
TT^{-1} = T^{-1}T = \one
  = \lrb \ba{rr} 1 & 0\\ 0 & 1 \ea \rrb\,.
\eeq
A matrix $T = \lrb \ba{rr} A & B \\ C & D \ea \rrb$ is called a $2 \times 2$ 
quantum  matrix if its matrix elements $\lcb A,B,C,D\rcb$ satisfy the 
commutation relations in~(\ref{Trel}).  In the limit $q \longrightarrow 1$ a 
quantum matrix $T$ becomes a classical matrix with commuting elements.  Note 
that the identity matrix $\one = \lrb \ba{rr} 1 & 0\\ 0 & 1 \ea \rrb$ is a 
quantum matrix.  

Entries of a quantum matrix $T$ are noncommuting variables satisfying the 
commutation relations~(\ref{Trel}).  Let 
$T_1 = \lrb \ba{rr} A_1 & B_1 \\ C_1 & D_1 \ea \rrb$ and 
$T_2 = \lrb \ba{rr} A_2 & B_2 \\ C_2 & D_2 \ea \rrb$ be any two quantum  
matrices; {\em i.e.}, $\lcb A_1,B_1,C_1,D_1 \rcb$ obey the 
relations~(\ref{Trel}), and $\lcb A_2,B_2,C_2,D_2 \rcb$ also obey the  
relations~(\ref{Trel}).  The matrix elements of $T_1$ and $T_2$ may be ordinary 
classical matrices satisfying the required relations~(\ref{Trel}).  Define the 
product 
\bea
\Delta_{12}(T) & = & T_1 \dot{\otimes} T_2 
  = \lrb \ba{rr} A_1 & B_1 \\ C_1 & D_1 \ea \rrb 
    \dot{\otimes} 
    \lrb \ba{rr} A_2 & B_2 \\ C_2 & D_2 \ea \rrb \nn \\ 
   & = & 
    \lrb \ba{rr} 
    A_1 \otimes A_2 + B_1 \otimes C_2 & 
    A_1 \otimes B_2 + B_1 \otimes D_2 \\ 
    C_1 \otimes A_2 + D_1 \otimes C_2 & 
    C_1 \otimes B_2 + D_1 \otimes D_2 
    \ea \rrb \nn \\
   & = & \lrb \ba{rr} 
         \Delta_{12}(A) & \Delta_{12}(B) \\ 
         \Delta_{12}(C) & \Delta_{12}(D) 
         \ea \rrb
\eea
where $\otimes$ denotes the direct product with the property 
$(P \otimes R)(Q \otimes S) = PQ \otimes RS$.  Then one finds that the matrix 
elements of $\Delta_{12}(T)$, namely, 
\bea
\Delta_{12}(A) & = & A_1 \otimes A_2 + B_1 \otimes C_2\,, \quad 
\Delta_{12}(B) = A_1 \otimes B_2 + B_1 \otimes D_2\,, \nn \\
\Delta_{12}(C) & = & C_1 \otimes A_2 + D_1 \otimes C_2\,, \quad 
\Delta_{12}(D) = C_1 \otimes B_2 + D_1 \otimes D_2\,, 
\label{delta}
\eea
also satisfy the commutation relations~(\ref{Trel}).  In other words, 
$\Delta_{12}(T)$ is also a quantum matrix.  This product, 
$\Delta_{12}(T) = T_1 \dot{\otimes} T_2$, is called the coproduct or 
comultiplication.  Note that there is no inverse for this coproduct.  Under 
this coproduct the $2 \times 2$ quantum matrices 
$\lcb T = \lrb \ba{rr} A & B \\ C & D \ea \rrb \rcb $ form a pseudomatrix 
group, commonly called a quantum group, denoted by $SL_q(2)$.  The algebra of 
functions over $SL_q(2)$, or the algebra of polynomials in $\lcb A,B,C,D\rcb$, 
is denoted by $Fun_q(SL(2))$. The coproduct operation ($\Delta$) 
defined by~(\ref{delta}), is symbolically written as 
\bea
\Delta(A) & = & A \otimes A + B \otimes C\,, \quad 
\Delta(B) = A \otimes B + B \otimes D\,, \nn \\
\Delta(C) & = & C \otimes A + D \otimes C\,, \quad 
\Delta(D) = C \otimes B + D \otimes D\,.  
\eea  
For any $f(A,B,C,D) \in Fun_q(SL(2))$ the definition of $\Delta$ is extended as 
$\Delta f(A,B,C,D) = f(\Delta(A),\Delta(B),\Delta(C),\Delta(D))$.  The algebra 
$Fun_q(SL(2))$ is technically a Hopf algebra.  To complete this algebraic 
structure two more operations called the coinverse (or antipode) denoted by 
$S$, and the counit denoted by $\varepsilon$, are defined and these operations 
$\Delta$, $S$, and $\varepsilon$ are required to satisfy certain axioms.  We 
shall not consider these details of the Hopf algebraic structure.  From the 
point of view physical applications quantum groups provide a generalization of 
symmetry concepts and involve mainly two fundamental new ideas\,: deformation 
and noncommutative  comultiplication. 

The matrix $T = \lrb \ba{rr} A & B \\ C & D \ea \rrb$ corresponds to the 
fundamental irreducible representation of $SL_q(2)$.  Higher dimensional 
representations are defined as follows. An $n \times n$ matrix 
\beq
T = \lrb T_{ij} \rrb 
  = \lrb \ba{cccc}
         T_{11} & T_{12} & \dots & T_{1n} \\
         T_{21} & T_{22} & \dots & T_{2n} \\
            .   &   .    & \dots &   .    \\
            .   &   .    & \dots &   .    \\
         T_{n1} & T_{n2} & \dots & T_{nn} 
         \ea \rrb\,, 
\eeq
is said to be an $n$-dimensional representation of $SL_q(2)$ if its matrix 
elements $\lrb T_{ij} \rrb$ are polynomials in $\lcb A,B,C,D\rcb$, or in other 
words elements of $Fun_q(SL(2))$, and satisfy the property 
\bea
\lrb T \dot{\otimes} T \rrb_{ij} & = & \sum_{l=1}^n T_{il} \otimes T_{lj} 
   = \Delta\lrb T_{ij} \rrb \nn \\
   & = & T_{ij}(\Delta(A),\Delta(B),\Delta(C),\Delta(D))\,, \quad  
         \forall\ i,j = 1,2,\dots, n\,.
\label{rep}
\eea  
Now, for example, look at 
\beq
T^{(1)} = \lrb \ba{ccc} 
          A^2 & \sqrt{1+q^{-2}}AB & B^2 \\
          \sqrt{1+q^{-2}}AC & AD+q^{-1}BC & \sqrt{1+q^{-2}}BD \\
          C^2 & \sqrt{1+q^{-2}}CD & D^2 
          \ea \rrb\,. 
\label{T1}
\eeq
It can be verified that $T^{(1)}$ is the $3$-dimensional representation of 
$SL_q(2)$.  For instance, see that 
\bea
\lrb T^{(1)} \dot{\otimes} T^{(1)} \rrb_{11} 
   & = & \sum_{l=1}^3 T^{(1)}_{1l} \otimes T^{(1)}_{l1} \nn \\
   & = & A^2 \otimes A^2 + \lrb 1+q^{-2} \rrb AB \otimes AC 
                         + B^2 \otimes C^2 \nn \\
   & = & A^2 \otimes A^2 + AB \otimes AC + BA \otimes CA 
                         + B^2 \otimes C^2 \nn \\          
   & = & (A \otimes A + B \otimes C)^2 = (\Delta(A))^2 
     = \Delta\lrb T^{(1)}_{11} \rrb\,, 
\eea 
as required.  Similarly, for other matrix elements of $T^{(1)}$ one can verify 
the property~(\ref{rep}), namely, 
\beq
\lrb T \dot{\otimes} T \rrb_{ij} = \Delta\lrb T_{ij}\rrb\,.
\eeq  

In the theory of classical Lie groups we know that an element of a Lie group 
$G$, say $g$, can be written as 
\beq
g = e^{\epsilon_1 L_1}e^{\epsilon_2 L_2}\cdots e^{\epsilon_n L_n}\,, 
\eeq
where the parameters $\lcb\epsilon_i\rcb$ characterize the group element $g$ 
and $\lcb L_i\rcb$ are constant generators of the group $G$ satisfying a Lie 
algebra 
\beq
\lsb L_i,L_j \rsb = \sum_{k=1}^n C_{ij}^k\,L_k\,, \quad 
                    i,j = 1,2,\dots,n\,, 
\eeq
with $\lcb C_{ij}^k\rcb$ as the structure constants.  When the group element 
$g$ is close to the identity element ($I$) of the group, the parameters 
$\lcb\epsilon_i\rcb$ are infinitesimals and one can write 
\beq
g \approx I + \sum_{i=1}^n \epsilon_i L_i\,.
\eeq 
Now, the interesting question is 
\begin{itemize}
\item Is there an analogue of the Lie algebra in the case of a quantum group? 
\end{itemize} 
The answer is yes!  To this end, first we have to recall some basic notions of 
the theory of $q$-series.  

One defines the $q$-shifted factorial by 
\beq
(x;q)_n = \lcb 
\ba{ll}
1, & n = 0, \\
(1-x)(1-xq)(1-xq^2) \dots\,\lrb 1-xq^{n-1} \rrb\,, & n = 1,2,\dots\,. 
\ea \right. 
\eeq
Then, with the notation
\beq
\lrb x_1,x_2,\,\dots\,,x_m ; q \rrb_n = 
\lrb x_1;q \rrb _n \lrb x_2;q \rrb _n\,\dots\,
   \lrb x_m;q \rrb_n \,, 
\eeq
an $_r\phi_s$ basic hypergeometric series, or a general $q$-hypergeometric 
series, is given by  
\bea
   &   & 
{}_r\phi_s \lrb a_1, a_2,\,\dots\,,a_r;b_1,
    b_2,\,\dots\,,b_s;q,z \rrb \nn \\ 
   &   & \quad = \sum_{n=0}^\infty\,\frac{
   \lrb a_1,a_2,\,\dots\,,a_r ; q \rrb_n}
{\lrb b_1,b_2,\,\dots\,,b_s ; q \rrb_n (q;q)_n} 
\lrb (-1)^n q^{n(n-1)/2} \rrb ^{1+s-r} z^n\,, \nn \\
   &   & \qquad \qquad \qquad \qquad \qquad \qquad \qquad \qquad  
r,s = 0,1,2,\dots\,. 
\label{phirs}
\eea
Consider 
\beq
{}_1\phi_0(0;-;q,(1-q)z) = e_q^z = 
          \sum_{n=0}^\infty\frac{z^n}{[n]_q!}\,,
\label{eqz}
\eeq
where 
\bea
   &   & [n]_q = \frac{1-q^n}{1-q}\,, \nn \\
   &   & [n]_q! = [n]_q[n-1]_q[n-2]_q\dots[2]_q[1]_q\,, \quad 
                    n = 1,2,\dots\,,\ \ [0]_q! = 1\,.
\eea 
The $q$-number $[n]_q$, or the so-called basic number, was defined by Heine 
(1846).  Much of Ramanujan's work is related to these $q$-series.  Note that 
\beq
[n]_q \stackrel{q \rightarrow 1}{\longrightarrow} n\,, \quad 
e_q^z \stackrel{q \rightarrow 1}{\longrightarrow} e^z\,. 
\eeq
Thus, $e_q^z$ defined by~(\ref{eqz}) is a $q$-generalization of the exponential 
function, called a $q$-exponential function\,; note that there can be several 
generalizations of the exponential function satisfying the condition that in 
the limit $q \rightarrow 1$ it should become the standard exponential function.  
In the theory of quantum groups a new definition of the $q$-number is often 
useful. It is 
\beq
\dlb n \drb_q = \frac{q^n-q^{-n}}{q-q^{-1}}\,.
\eeq 
Note that $\dlb n \drb_q $ also becomes $n$ in the limit $q \rightarrow 1$, and 
$\dlb n \drb_q $ is symmetric with respect to the interchange of $q$ and $q^{-1}$ 
unlike Heine's $[n]_q$.  

Now consider the $2$-dimensional $T$-matrix parametrized as 
\beq
T = \lrb \ba{rr} A & B \\ C & D \ea \rrb 
  = \lrb \ba{cc} 
         e^\alpha & e^\alpha\beta \\
         \gamma e^\alpha & e^{-\alpha}+\gamma e^\alpha\beta 
         \ea \rrb\,, 
\label{albega}
\eeq
which requires the variable parameters $\lcb\alpha, \beta, \gamma\rcb$ to  
satisfy a Lie algebra 
\beq [\alpha,\beta] = (\log q)\,\beta\,, \quad 
[\alpha,\gamma] = (\log q)\,\gamma\,, \quad 
[\beta,\gamma] = 0\,, 
\eeq
so that $\lcb A,B,C,D\rcb$ obey the algebra~(\ref{Trel}).  Then, one can write 
\beq
T = e_{q^{-2}}^{\gamma {\cal X}_-^{(1/2)}} e^{2\alpha {\cal X}_0^{(1/2)}}  
        e_{q^2}^{\beta {\cal X}_+^{(1/2)}}\,, 
\label{Trep}
\eeq
with 
\beq
{\cal X}_0^{(1/2)} = \frac{1}{2}\lrb\ba{rr} 1 & 0 \\ 0 & -1 \ea\rrb\,, 
              \quad 
{\cal X}_-^{(1/2)} = \lrb\ba{rr} 0 & 0 \\ 1 & 0 \ea\rrb\,, \quad  
{\cal X}_+^{(1/2)} = \lrb\ba{rr} 0 & 1 \\ 0 & 0 \ea\rrb\,. 
\label{J1/2}
\eeq
Of course, in this case, $e_{q^{-2}}^{\gamma {\cal X}_-^{(1/2)}}$ and   
$e_{q^2}^{\beta {\cal X}_+^{(1/2)}}$ are trivially the same as 
$e^{\gamma {\cal X}_-^{(1/2)}}$ and $e^{\beta {\cal X}_+^{(1/2)}}$, respectively,  
since $\lrb {\cal X}_\pm^{(1/2)}\rrb^2 = 0$.  

Actually, equation~(\ref{Trep}) is the special case of a universal formula for 
the representations of the $T$-matrices of $SL_q(2)$ and corresponds to the 
fundamental representation.  The generic form of the $T$-matrix is given by 
\beq
T = e_{q^{-2}}^{\gamma {\cal X}_-} e^{2\alpha {\cal X}_0} 
e_{q^2}^{\beta{\cal X}_+}\,, 
\label{univT}
\eeq
called the universal ${\cal T}$-matrix, where 
$\lcb {\cal X}_0, {\cal X}_+, {\cal X}_-\rcb$ obey the algebra 
\bea
\lsb {\cal X}_0\,,\,{\cal X}_\pm \rsb = \pm {\cal X}_\pm\,, \quad 
\lsb {\cal X}_+\,,\,{\cal X}_- \rsb 
= \frac{q^{2{\cal X}_0}-q^{-2{\cal X}_0}}{q - q^{-1}} 
= \dlb 2{\cal X}_0 \drb_q\,, 
\label{slq2}
\eea
called the quantum algebra $sl_q(2)$.  In the limit $q \rightarrow 1$,  
$\{{\cal X}\} \rightarrow \{X\}$ which generate $sl(2)$, 
\beq
\lsb X_0\,,\,X_\pm \rsb = \pm X_\pm\,, \quad 
\lsb X_+\,,\,X_-\rsb = 2X_0\,, 
\label{sl2}
\eeq
the Lie algebra of $SL(2)$. The matrices 
$\lcb {\cal X}_0^{(1/2)},{\cal X}_+^{(1/2)},{\cal X}_-^{(1/2)}\rcb$  
in~(\ref{J1/2}) provide the fundamental $2$-dimensional irreducible 
representation of the algebra~(\ref{slq2}) (actually, they also provide the 
fundamental representation of the generators of $sl(2)$ algebra~(\ref{sl2})).  
When a higher dimensional representation of~(\ref{slq2}) is plugged in the 
formula~(\ref{univT}) ${\cal T}$ becomes a higher dimensional representation of 
the $T$-matrix.  For example, the three dimensional representation~(\ref{T1}) 
is obtained by substituting in~(\ref{univT}) 
\bea
{\cal X}_0^{(1)} & = & \lrb\ba{rrr} 1 & 0 &  0 \\ 
                             0 & 0 &  0 \\ 
                             0 & 0 & -1 \ea\rrb\,, \nn \\ 
{\cal X}_-^{(1)} & = & \sqrt{\dlb 2\drb_q}\lrb\ba{ccc} 
                             0 & 0 &  0 \\ 
                             \sqrt{q} & 0 & 0 \\
                             0 & 1/\sqrt{q} & 0 \ea\rrb\,, \nn \\ 
{\cal X}_+^{(1)} & = & \sqrt{\dlb 2\drb_q}\lrb\ba{ccc} 
                             0 & 1/\sqrt{q} & 0 \\
                             0 & 0 & \sqrt{q} \\
                             0 & 0 & 0 \ea\rrb\,,
\label{J1}
\eea
which provide the three dimensional irreducible representation of the $sl_q(2)$ 
algebra~(\ref{slq2}), and using the parametrization of $\lcb A,B,C,D\rcb$ in 
terms of $\lcb\alpha, \beta, \gamma\rcb$ as given by~(\ref{albega}).  Note that 
in the limit $q \rightarrow 1$ these matrices~(\ref{J1}) obey the $sl(2)$ 
algebra~(\ref{sl2}).  As seen from~(\ref{albega}), (\ref{univT}) and~(\ref{slq2}), 
one can say that for a quantum group the group-parameter space is noncommutative.  

The algebra $sl_q(2)$ is also often called a quantum group.  Actually, the   
relations in~(\ref{slq2}) define the generators of the $q$-deformation of the 
universal enveloping algebra of $sl(2)$.  Hence, the relations~(\ref{slq2}) are 
also referred to, more properly, as $U_q(sl(2))$,  the $q$-deformed universal 
enveloping algebra of $sl(2)$.  The algebra $U_q(sl(2))$, generated by 
polynomials in $\lcb {\cal X}_0, {\cal X}_+, {\cal X}_-\rcb$ obeying the 
relations~(\ref{slq2}), is also a Hopf algebra.  We shall consider only the 
coproduct(s) for $U_q(sl(2))$.  A coproduct for $sl_q(2)$ is  
\bea
\Delta_q\lrb {\cal X}_0\rrb = 
{\cal X}_0 \otimes \one + \one \otimes {\cal X}_0\,, \quad  
\Delta_q\lrb {\cal X}_\pm \rrb = {\cal X}_\pm \otimes q^{{\cal X}_0} 
               + q^{-{\cal X}_0} \otimes {\cal X}_\pm\,. 
\label{qcomu}
\eea
It can be easily verified that this comultiplication rule is an algebra  
isomorphism for $sl_q(2)$\,: 
\bea
\lsb \Delta_q\lrb {\cal X}_0\rrb, \Delta_q\lrb {\cal X}_\pm\rrb\rsb 
                    = \pm\Delta_q\lrb {\cal X}_\pm\rrb\,, \quad   
\lsb \Delta_q\lrb {\cal X}_+\rrb, \Delta_q\lrb {\cal X}_-\rrb\rsb 
                    = \dlb 2\Delta_q\lrb {\cal X}_0\rrb\drb_q\,.
\eea
The most important property of this coproduct is its noncommutativity.  Note 
that the algebra~(\ref{slq2}) is invariant under the interchange 
$q \leftrightarrow q^{-1}$ since 
$\dlb 2{\cal X}_0 \drb_q = \dlb 2{\cal X}_0 \drb_{q^{-1}}$.  However, 
the comultiplication~(\ref{qcomu}) is not invariant under such an interchange.  
This means that the comultiplication obtained from~(\ref{qcomu}) by an 
interchange $q \leftrightarrow q^{-1}$ should also be an equally good 
comultiplication.  It can be verified that the coproduct so obtained, namely,  
\bea
\Delta_{q^{-1}}\lrb {\cal X}_0\rrb 
    = {\cal X}_0 \otimes \one + \one \otimes {\cal X}_0\,, \quad  
\Delta_{q^{-1}}\lrb {\cal X}_\pm \rrb 
    = {\cal X}_\pm \otimes q^{-{\cal X}_0} 
              + q^{{\cal X}_0} \otimes {\cal X}_\pm\,. 
\label{q1comu}
\eea
is indeed an algebra isomorphism for $sl_q(2)$.  This coproduct~(\ref{q1comu}), 
$\Delta_{q^{-1}}$, is called the opposite coproduct in view of the relation 
\beq
\Delta_{q^{-1}}({\cal X}) = \tau\lrb\Delta_q({\cal X})\rrb\,,\ \ 
             \mbox{where}\ \ \tau(u\otimes v) = v\otimes u\,.
\eeq
Since 
\beq
\Delta_{q^{-1}} \neq \Delta_q\,, \quad 
\mbox{or}\ \ \tau(\Delta) \neq \Delta \,, 
\eeq
the comultiplications $\Delta_q$ and $\Delta_{q^{-1}}$ of $sl_q(2)$ are 
noncommutative.  In the limit $q \rightarrow 1$, the classical $sl(2)$ has only 
a single  comultiplication, $\Delta(X) = X \otimes \one + \one \otimes X$\,, 
which is commutative ({\em i.e.}, $\tau(\Delta) = \Delta$).  

One can show that the two comultiplications of $sl_2(q)$, namely $\Delta_q$ and 
$\Delta_{q^{-1}}$, are related to each other by an equivalence relation such 
that there exists an ${\cal R} \in U_q(sl(2)) \otimes U_q(sl(2))$, called the 
universal ${\cal R}$-matrix, satisfying the relation 
\beq
\Delta_{q^{-1}}({\cal X}) = {\cal R}\Delta_q({\cal X}){\cal R}^{-1}\,.
\eeq
This universal ${\cal R}$-matrix is the central object of the quantum group 
theory.  In this case it can be shown that  
\beq
{\cal R} = q^{2\lrb {\cal X}_0 \otimes {\cal X}_0 \rrb} 
           \sum_{n=0}^\infty 
           \frac{\lrb 1-q^2\rrb^n}{\dlb n\drb_q!} 
           q^{n(n-1)/2}\lrb q^{{\cal X}_0}{\cal X}_+ 
           \otimes q^{-{\cal X}_0}{\cal X}_-\rrb^n\,.
\label{univR}
\eeq
If we insert the matrix representations of $\{{\cal X}\}$ in this expression 
for ${\cal R}$ we get numerical $R$-matrices.  For example, substituting 
in~(\ref{univR}) the $2\times 2$ representation of $\{{\cal X}\}$, given 
in~(\ref{J1/2}), we get the fundamental $4$-dimensional $R$-matrix 
\beq
R = \frac{1}{\sqrt{q}}
    \lrb\ba{cccc}
    q & 0 & 0 & 0 \\
    0 & 1 & \lrb q-q^{-1}\rrb & 0 \\
    0 & 0 & 1 & 0 \\
    0 & 0 & 0 & 1 
    \ea\rrb\,.
\label{R}
\eeq
Let us now write any $R$ in the form 
\beq
R = \sum_i a_i\,u_i\otimes v_i\,. 
\eeq
It is clear from~(\ref{univR}) that this can be done.  Now, define
\bea
R_{12} = R \otimes \one\,, \quad    
R_{13} = \sum_i a_i\,u_i\otimes \one \otimes v_i\,, \quad   
R_{23} = \one \otimes R\,. 
\eea
Then, these satisfy the remarkable relation 
\beq
R_{12}R_{13} R_{23} = R_{23}R_{13} R_{12}\,. 
\label{ybe}
\eeq
known as the quantum Yang-Baxter equation, or simply the Yang-Baxter  
equation~(YBE). 

We have considered only the simplest example of a quantum group, namely  
$SL_q(2)$, associated with the classical group $SL(2)$.  There exists a  
systematic theory of deformation of any classical group.  It is also possible, 
in certain cases, to obtain deformations with several $q$-parameters.  Actually, 
the study of quantum groups sheds more light on the structure of the classical 
group theory.  I shall not go further into the details of the formalism of 
quantum group theory.  

Now, I am in a position to mention a few applications of quantum groups and 
algebras.  First, let us see how these things started.  Define 
\bea
T_1 & = & T \otimes \one 
    = \lrb\ba{cc} A & B \\ C & D \ea\rrb 
      \otimes \lrb\ba{cc} 1 & 0 \\ 0 & 1 \ea\rrb\,, \nn \\
T_2 & = & \one \otimes T  
    = \lrb\ba{cc} 1 & 0 \\ 0 & 1 \ea\rrb 
      \otimes \lrb\ba{cc} A & B \\ C & D \ea\rrb\,. 
\label{t1t2}
\eea 
Note that 
\beq
T_1T_2 = \lrb\ba{cccc}
         A^2 & AB & BA & B^2 \\
         AC  & AD & BC & BD \\
         CA  & CB & DA & DB \\
         C^2 & CD & DC & D^2 
         \ea\rrb \neq 
T_2T_1 = \lrb\ba{cccc}
         A^2 & BA & AB & B^2 \\
         CA  & DA & CB & DB \\
         AC  & BC & AD & BD \\
         C^2 & DC & CD & D^2 
         \ea\rrb\,, 
\eeq
because $\lcb A,B,C,D\rcb$ ae noncommutative.  The relation between $T_1T_2$ and 
$T_2T_1$ turns out to be  
\beq
RT_1T_2 = T_2T_1R\,.   
\label{rtt}
\eeq
This type of relation is commonly encountered in the quantum inverse scattering 
method approach to integrable models in quantum field theory and statistical 
mechanics.  Substituting in~(\ref{rtt}) $R$ from~(\ref{R}), and $T_1$ and 
$T_2$ from~(\ref{t1t2}), it is found that equation~(\ref{rtt}) is a 
compact way of stating the commutation relations~(\ref{Trel}) defining 
the fundamental $T$-matrix of $SL_q(2)$.  What about the commutation 
relations~(\ref{slq2}) defining $sl_q(2)$?   Define 
\bea
L^{(+)} & = & \lrb\ba{cc}
          q^{-{\cal X}_0} & -\sqrt{q}\lrb q-q^{-1}\rrb {\cal X}_- \\
                 0        & q^{{\cal X}_0} 
          \ea\rrb\,, \nn \\  
L^{(-)} & = & \lrb\ba{cc}
          q^{{\cal X}_0}  &      0        \\
          q^{-1/2}\lrb q-q^{-1}\rrb {\cal X}_+ & q^{-{\cal X}_0} 
          \ea\rrb\,, \nn \\ 
L^{(\pm)}_1 &  = & L^{(\pm)} \otimes \one\,, \quad 
L^{(\pm)}_2 = \one \otimes L^{(\pm)}\,.
\eea 
Then, the commutation relations~(\ref{slq2}), defining the generators of   
$sl_q(2)$, can be stated elegantly as \bea
R^{-1}L^{(\pm)}_1L^{(\pm)}_2 = L^{(\pm)}_2L^{(\pm)}_1R^{-1}\,, \quad 
R^{-1}L^{(+)}_1L^{(-)}_2 = L^{(-)}_2L^{(+)}_1R^{-1}\,. 
\eea
Note that the $L^{(\pm)}$-matrices are special realizations of the  
$T$-matrices, {\em i.e.}, the elements of $L^{(\pm)}$-matrices obey the 
commutation relations required of the $T$-matrix elements.  

If we define for the $R$-matrix in~(\ref{R}), 
\beq
S_1 = \check{R} \otimes \one, \quad 
S_2 = \one \otimes \check{R}\,, 
\eeq
where
\beq 
\check{R} = PR\,, \quad 
P = \lrb\ba{cccc}
1 & 0 & 0 & 0 \\
0 & 0 & 1 & 0 \\
0 & 1 & 0 & 0 \\
0 & 0 & 0 & 1 
\ea\rrb\,, 
\eeq
then, it is found that 
\beq
S_1S_2S_1 = S_2S_1S_2\,, 
\label{s121}
\eeq
which is an alternative form of the YBE~(\ref{ybe}).  For any general 
$R$-matrix the YBE~(\ref{ybe}) can be put in this form~(\ref{s121}).  This 
relation~(\ref{s121}) represents a property of the generators of a braid group 
which is a generalization of the well known symmetric group $S_n$.  The 
symmetric group $S_n$ is the group of all permutations of $n$ given objects.  
An element of the braid group $B_n$ can be depicted as a system of $n$ strings 
joining two sets of $n$ points, each set located on a line, the two lines, say 
top and bottom, being parallel, with over-crossings or under-crossings of the 
strings.  The over-crossings and the under-crossings of the strings make $B_n$ 
an infinite group which will otherwise reduce to $S_n$. If $i$ and $i+1$ are two  
consecutive points on the top and bottom lines, the string starting at $i$ on 
the top line can reach $i+1$ on the bottom line by either under-crossing or 
over-crossing the string starting at $i+1$ on the top line and reaching $i$ on 
the bottom line.  The corresponding elements of the braid group are usually 
denoted by $\sigma_i$ and $\sigma_i^{-1}$, respectively.  The elements 
$\lcb \sigma_i \left|\,i = 1,2,\cdots\,n-1\right. \rcb $\,, generating the 
braid group $B_n$, satisfy two relations, 
\beq
\sigma_i\sigma_j = \sigma_j\sigma_i \quad 
\mbox{\rm for}\ \ |i-j| > 1\,, 
\eeq
and 
\beq
\sigma_i\sigma_{i+1}\sigma_i = \sigma_{i+1}\sigma_i\sigma_{i+1}\,.
\label{braid}
\eeq 
Now, comparing the relations~(\ref{s121}) and~(\ref{braid}) it is obvious that 
the solutions of the YBE ($R$-matrices), or the quantum algebras, should play a 
central role in the theory of representations of braid groups.  Braid groups 
have many applications.  In mathematics they are useful in the study of complex 
functions of hypergeometric type having several variables.  In physics they 
appear in knot theory, statistical mechanics, two-dimensional conformal field 
theory, and so on.  

In quantum mechanics the notion of continuous space-time with commutative 
coordinates is taken over from classical mechanics.  This is an assumption.  
What will happen if at some deeper microscopic level the space-time coordinates 
themselves are noncommutative?  It is clear that to deal with such a situation 
one will have to use a noncommutative differential calculus and the theory of 
quantum groups provides the necessary framework as we have seen above.  For 
example, consider the motion of a quantum particle in a two dimensional 
noncommutative plane with $X$ and $Y$ as the coordinates.  If we take the 
corresponding conjugate momenta to be proportional to $\frac{\pl}{\pl X}$ and 
$\frac{\pl}{\pl Y}$, respectively, then the relations~(\ref{XYrel}) indicate 
how the two-dimensional quantum mechanical phase-space would be deformed at 
that level.  Thus, the theory of quantum groups would provide the mathematical 
framework for the future of quantum physics if it turns out that at some deeper 
microscopic level space-time manifold is noncommutative.  Hence there has been 
a lot of interest in studying the fundamental modifications that would occur in 
the framework quantum mechanics, relativity theory, Poincar\'{e} group, ..., 
etc., if the space-time manifold happens to be noncommutative.   

Apart from applications of fundamental nature, such as those mentioned above, 
there have been many phenomenological applications of quantum algebras in 
nuclear physics, condensed matter physics, molecular physics, quantum optics, 
and elementary particle physics.  In these applications either an existing 
model is identified with a quantum algebraic structure, or a standard model is 
deformed to have an underlying quantum algebraic structure and studied to 
reveal the new features emerging.  To give an idea of such applications, let me 
mention, as the final example, the $q$-deformation of the quantum mechanical 
harmonic oscillator algebra, also known as the boson algebra.  The algebraic 
treatment of the quantum mechanical harmonic oscillator involves a creation 
operator  $\left(a^\dagger \right)$, an annihilation operator $(a)$, and a 
number operator $(N)$, obeying the commutation relations 
\beq
\lsb a\,,\,a^\dagger\rsb = 1\,, \quad 
\lsb N\,,\,a^\dagger\rsb = a^\dagger\,, 
\label{bose}
\eeq
where $N$ is a hermitian operator and $a^\dagger$ is the hermitian conjugate of 
$a$.  The energy spectrum of the harmonic oscillator is given by the 
eigenvalues of the Hamiltonian operator 
\beq
H = \frac{1}{2}\left(aa^\dagger + a^\dagger a\right)\,, 
\eeq
in the appropriate units.  Taking two such sets of oscillator operators, 
$\lcb a_1, a_1^\dagger, N_1\rcb$ and $\lcb a_2, a_2^\dagger, N_2\rcb$, which 
are assumed to commute with each other, and defining 
\beq
J_0 = \frac{1}{2}\left(N_1 - N_2\right)\,, \quad
J_+ = a_1^\dagger a_2\,, \quad
J_- = a_2^\dagger a_1\,, 
\label{js}
\eeq
it is found that 
\beq
J_0^\dagger = J_0\,, \quad J_+^\dagger = J_-\,, 
\label{hc}
\eeq
and 
\beq
\lsb J_0\,,\,J_\pm\rsb = \pm J_\pm\,, \quad 
\lsb J_+\,,\,J_-\rsb = 2J_0\,.
\label{su2}
\eeq
This Lie algebra~(\ref{su2}) is seen to be the same as the $sl(2)$ 
algebra~(\ref{sl2}) subject to the hermiticity conditions~(\ref{hc}) and is 
known as $su(2)$ algebra, the Lie algebra of the group $SU(2)$.  The $su(2)$ 
algebra is the algebra of three dimensional rotations, or the rigid rotator, 
with $\lcb J_0, J_\pm\rcb$ representing the angular momentum operators.  The 
coproduct rule 
\beq
\Delta\lrb J_0 \rrb = J_0 \otimes \one + \one \otimes J_0\,, \quad
\Delta\lrb J_\pm \rrb = J_\pm \otimes \one + \one \otimes J_\pm\,, 
\eeq
for the algebra~(\ref{su2}), obtained by setting $q = 1$ in~(\ref{qcomu}) 
(or~(\ref{q1comu})), represents the rule for addition of angular momenta.  
Correspondingly, the relations~(\ref{slq2}) rewritten as 
\beq
\lsb{\cal J}_0\,,\,{\cal J}_\pm\rsb = \pm {\cal J}_\pm\,, \quad
\lsb{\cal J}_+\,,\,{\cal J}_-\rsb = \dlb 2J_0 \drb_q\,, 
\eeq 
with the hermiticity conditions 
\beq
{\cal J}_0^\dagger = {\cal J}_0\,, \quad
{\cal J}_+^\dagger = {\cal J}_-\,, 
\eeq
represent the $su_q(2)$ (or $U_q(su(2))$) algebra or the $q$-deformed version 
of the $su(2)$ algebra~(\ref{su2}).  One can say that $su_q(2)$ is the algebra 
of the $q$-rotator.  For the $q$-angular momentum operators there are two 
possible addition rules,  
\beq
\Delta_{q^{\pm 1}}\lrb{\cal J}_0\rrb =  
{\cal J}_0 \otimes \one + \one \otimes {\cal J}_0\,, \quad  
\Delta_{q^{\pm 1}}\lrb{\cal J}_\pm\rrb = {\cal J}_\pm \otimes q^{\pm{\cal J}_0} 
               + q^{\mp{\cal J}_0} \otimes {\cal J}_\pm\,, 
\eeq
as seen from~(\ref{qcomu}) and~(\ref{q1comu}).  Now the interesting fact is that 
one has a realization of $su_q(2)$ generators given by
\beq
{\cal J}_0 = \frac{1}{2}\left({\cal N}_1 - {\cal N}_2\right)\,, \quad
{\cal J}_+ = A_1^\dagger A_2\,, \quad
{\cal J}_- = A_2^\dagger A_1\,, 
\eeq
exactly analogous to the $su(2)$ case~(\ref{js}), where the two sets of 
operators $\lcb A_1, A_1^\dagger, {\cal N}_1\rcb$ and  
$\lcb A_2, A_2^\dagger, {\cal N}_2\rcb$ commute with each other and obey, 
within each set, the algebra 
\beq
AA^\dagger - qA^\dagger A = q^{-{\cal N}}\,, \quad 
\lsb {\cal N}\,,\,A^\dagger\rsb = A^\dagger\,. 
\label{qbose}
\eeq
Further, ${\cal N}$ is hermitian and $\lcb A\,,\,A^\dagger \rcb$ is a 
hermitian conjugate pair.  The $q$-deformed oscillator algebra~(\ref{qbose}) 
is known as the $q$-oscillator or the $q$-boson algebra.  When 
$q \longrightarrow 1$ the $q$-oscillator algebra~(\ref{qbose}) reduces to the 
canonical oscillator algebra~(\ref{bose}).  As is easy to guess,  
phenomenological applications of quantum algebras in nuclear and molecular 
spectroscopy involve the substitution of harmonic oscillator model by the 
$q$-oscillator model and the rigid rotator model by the $q$-rotator model.  
Such applications lead to impressive results showing that the actual 
vibrational-rotational spectra of nuclei and molecules can be fit into schemes 
in which the number of phenomenological $q$-parameters required are very much 
fewer than the number of traditional phenomenological parameters required to 
fit the same spectral data.  Somehow such $q$-deformed models seem to take 
into account more efficiently the anharmonicity of vibrations and the 
nonrigidity of rotations in nuclear and molecular systems.  \\ 

\noindent
I wish to thank the Management, and Prof. Dr. K. V. Parthasarathy (Head, 
Department of Mathematics), of the Ramakrishna Mission Vivekananda College,  
for giving me the  privilege of delivering the {\em Narayani, N. and 
Kadayam S. Sankaran Memorial Lectures} for 1999.

\end{document}